# Quantum-inspired Huffman Coding

**A. S. Tolba, M. Z. Rashad, and M. A. El-Dosuky**

Dept. of Computer Science, Faculty of Computers and Information Sciences,
Mansoura University, Mansoura, Egypt.

tolba_1954@yahoo.com, magdi_12003@yahoo.com, dr_dos_ok@yahoo.com

July 2007

## ABSTRACT

Huffman Compression, also known as Huffman Coding, is one of many compression techniques in use today. The two important features of Huffman coding are *instantaneousness* that is the codes can be interpreted as soon as they are received and *variable length* that is a most frequent symbol has length smaller than a less frequent symbol. The traditional Huffman coding has two procedures: constructing a tree in $O(n^2)$ and then traversing it in $O(n)$.

Quantum computing is a promising approach of computation that is based on equations from Quantum Mechanics. Instantaneousness and variable length features are difficult to generalize to the quantum case. The quantum coding field is pioneered by Schumacher works on block coding scheme. To encode N signals *sequentially*, it requires $O(N^3)$ computational steps. The encoding and decoding processes are far from instantaneous. Moreover, the lengths of all the codewords are the same. A Huffman-coding-inspired scheme for the storage of quantum information takes $O(N(\log N)^a)$ computational steps for a sequential implementation on non-parallel machines.

The proposed algorithm, Quantum-inspired Huffman coding of symbols with equal frequencies, also has two procedures : calculating a quantum system state in $O(n (\lg n)^2)$ and then multiplying it by the inputs in $O((\lg n)^2)$. Finally, we present an unprecedented scheme for direct mapped Huffman codes that is $O(\lg n)$.

## 1. Introducing Huffman Coding

Information theory answers two fundamental questions in communication theory: what is the ultimate data compression (answer is the *entropy H*), and what is the ultimate transmission rate of communication (answer is the *channel capacity C*) [1]. Here, we are interested in the ultimate data compression. We shall begin by stating some common definitions. For more detailed illustrations, examples and exercises please refer to [1].

The *entropy* of a random variable $X$ with a probability mass function $p(x)$ is defined by
$$H(X) = - \sum p(x) \lg p(x)$$
where $\lg a$ means $\log_2 a$. The entropy is the number of bits on average required to describe the random variable. This establishes the entropy as a nature measure of efficient description length for words in a code.

A *source code* $C$ for a random variable $X$ is a mapping from $R$, the range of $X$, to $D^*$, the set of finite length strings of symbols from an $n$-ary alphabet $D = \{x_0, x_1, x_2, ..., x_{n-1}\}$. The alphabet D is usually shorthanded as $D=\{0, 1, ..., n-1\}$. Let $c(x)$ denotes the code word corresponding to $x$ and $l(x)$ denotes the length of $c(x)$.

The *expected length* $L(C)$ of a source code $C$ for a random variable $X$ with probability mass function $p(x)$ is given by
$$L(C) = \sum p(x) l(x)$$

Now, let $w_l$ denotes a word of length $l$ and formed as a sequence of $l$ symbols $(x_1, x_2, ..., x_l)$. A code is said to be *non-singular* if every element of the range of $X$ maps into a different string in $D^*$, i.e.,
$$x_i \neq x_j \Rightarrow c(x_i) \neq c(x_j)$$

The *extension* $C^*$ of a code $C$ is the mapping from finite length strings of $R$ to finite length strings of $D$, defined by
$$c(x_1, x_2, ..., x_l) = c(x_1)c(x_2)...c(x_l)$$
where $c(x_1)c(x_2)...c(x_l)$ indicates concatenation of the corresponding words.

A code is called *uniquely decodable* if its extension is non-singular. In other words, there is no ambiguity and we ensure decodability because any encoded string has only one possible source string producing it.

A code is called a *prefix code* or an *instantaneous code* if no codeword is a prefix of any other codeword. An instantaneous code can be interpreted codeword by codeword, as soon as they are received.

The following figure summarizes the relation between the classes of codes mentioned so far.

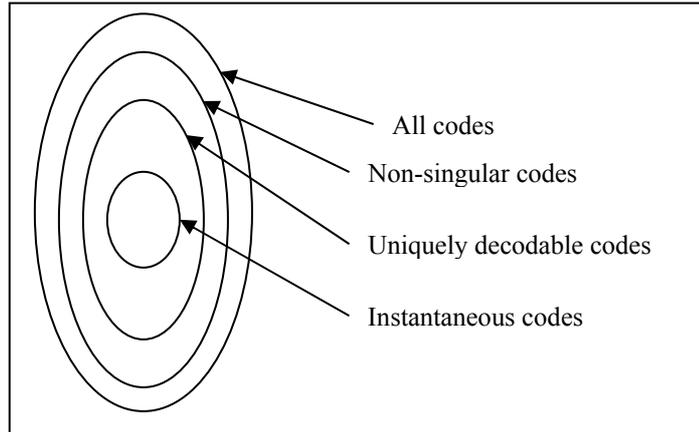

(Figure 1: Classes of Codes)

**Huffman Coding**

Huffman Compression, also known as Huffman Coding, is one of many compression techniques in use today. It is too easy to understand and implement yet still gets a decent compression ratio on average files.

The Huffman algorithm begins with a set of symbols each with its frequency of occurrence (probability) constructing what we can call a frequency table. The Huffman algorithm then builds the Huffman Tree using this frequency table.

The tree structure contains nodes, each contains a symbol, its frequency, a pointer to a parent node, and pointers to the left and right child nodes. The tree grows by making successive passes through the existing nodes. Each pass searches for two nodes that have the two lowest frequency counts, provided that they have not grown a parent node. When the algorithm finds those two nodes, it generate a new node, assigns it as the parent of the two nodes, and gives the new node a frequency count that is the sum of the two child nodes. The next iterations ignores those two child nodes but includes the new parent node. The passes continue until only one node with no parent remains. That node will be the root node of the tree.

Compression involves traversing the tree beginning at the leaf node for the symbol to be compressed and navigating to the root. This navigation iteratively selects the parent of the current node and sees whether the current node is the "right" or "left" child of the parent, thus determining if the next bit is a (1) or a (0). The final bit string is now to be reversed, because we are proceeding from leaf to root.

**Example:** Here is an example on Huffman coding, adapted from [1].
Let $X =$ { a, b, c, d, e } and $P(X) =$ { 0.25, 0.25, 0.2, 0.15, 0.15 }. The final codes are {00, 10, 11, 010, 011}

| $x$ | step 1 | step 2 | step 3 | step 4 |   | $a_i$ | $p_i$ | $h(p_i)$ | $l_i$ | $c(a_i)$ |
|---|---|---|---|---|---|---|---|---|---|---|
| a | 0.25 | 0.25 | 0.25 — 0 | 0.55 — 0 | 1.0 | a | 0.25 | 2.0 | 2 | 00 |
| b | 0.25 | 0.25 — 0 | 0.45 | 0.45  1 |   | b | 0.25 | 2.0 | 2 | 10 |
| c | 0.2 | 0.2  1 |   |   |   | c | 0.2 | 2.3 | 2 | 11 |
| d | 0.15 — 0 | 0.3 | 0.3  1 |   |   | d | 0.15 | 2.7 | 3 | 010 |
| e | 0.15  1 |   |   |   |   | e | 0.15 | 2.7 | 3 | 011 |

(Figure 2: Calculating Huffman Codes using a tree)

**Optimality of Huffman coding**

A worthy saying lemma presented and proved in [1], yet we state it here without proof. It states that for any distribution, there exists an optimal instantaneous code (with minimum expected length) that satisfies the following properties:
1. If $p_j > p_k$, then $l_j \leq l_k$, i.e. a most frequent symbol has length smaller than a less frequent symbol.
2. The two longest codewords have the same length
3. The two longest codewords differ only in the last bit and correspond to the two least likely symbols.

# 2. Introducing Quantum Computing

Quantum computing is a promising approach of computation that is based on equations from Quantum Mechanics. A Bit is the basic computational unit of computing. It encodes precisely one value, a 0 or a 1. A register of *n* bits can store **ANY** *n*-bit number. A *qubit* (or quantum *bit*) is the fundamental unit of quantum computing. A quantum bit exists in a superposition of states, and encodes the values 1 and 0 simultaneously. A quantum register of *n* qubits stores **ALL** *n*-bit numbers, i.e. $2^n$ values. Quantum computers use *Superposition* to allow qubits to store two different values at the same time [2].

**Quantum State**

The quantum state $|\psi\rangle$ represents a qubit if there are α, β ∈ C, where C is the set of Complex numbers, such that

$$|\psi\rangle = \alpha|0\rangle + \beta|1\rangle \quad \text{or}$$
$$|\psi\rangle = \sin\theta\,|0\rangle + \cos\theta\,|1\rangle$$

With $|\alpha|^2 + |\beta|^2 = 1$. $|0\rangle$ and $|1\rangle$ are the computational basis states. Measuring the state $|\psi\rangle$ with respect to $\{|0\rangle, |1\rangle\}$ basis will give $|0\rangle$ with probability $|\alpha|^2$ and $|1\rangle$ with probability $|\beta|^2$

The states $|+\rangle$ and $|-\rangle$ defined as

$$|+\rangle = \frac{1}{\sqrt{2}}|0\rangle + \frac{1}{\sqrt{2}}|1\rangle$$
$$|-\rangle = \frac{1}{\sqrt{2}}|0\rangle - \frac{1}{\sqrt{2}}|1\rangle$$

The states $|+\rangle$ and $|-\rangle$ form a basis. Note that:

$$|0\rangle = \frac{1}{\sqrt{2}}(|+\rangle + |-\rangle)$$
$$|1\rangle = \frac{1}{\sqrt{2}}(|+\rangle - |-\rangle)$$

**The interpretation of Qubits**

A bit represents one of two points. A qubit represents any point on the unit circle in the complex plane[3].

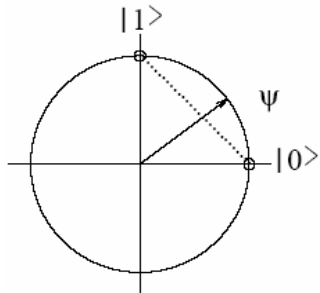

(Figure 3: A qubit as a point on the unit circle in the complex plane)

To ask the state of the qubit is to ask whether it is $|0\rangle$ or $|1\rangle$. Therefore, when we measure a qubit, we can only ever get $|0\rangle$ or $|1\rangle$, corresponding to Boolean 0 or 1. *Until we ask, it can be an arbitrary mixture of $|0\rangle$ and $|1\rangle$*.

**Matrix notation**

A 2-level quantum system can store a single qubit state. We will have

$$|0\rangle = \begin{pmatrix} 1 \\ 0 \end{pmatrix} \quad |1\rangle = \begin{pmatrix} 0 \\ 1 \end{pmatrix}$$

So,

$$a|0\rangle + b|1\rangle = \begin{pmatrix} a \\ b \end{pmatrix}$$

Also, we can say that: $|0\rangle = 01$ and $|1\rangle = 10$. That is $|x\rangle$ = binary (x + 1).

The symbol $|.\rangle$ is called a *ket*.

$$|\psi\rangle = \begin{pmatrix} z1 \\ \vdots \\ zn \end{pmatrix}$$

where $z1, \ldots, zn \in C$

The symbol ⟨.| is called a **bra**. ⟨ψ| represents the conjugate transpose of |ψ⟩.

$$\langle \psi | = \begin{pmatrix} z1 \\ \vdots \\ zn \end{pmatrix}^t = (\overline{z1}, \ldots, \overline{zn})$$

Writing |1⟩⟨0| + |0⟩⟨1| means mapping |1⟩ to ⟨0| and |0⟩ to ⟨1|. Note that |ψ⟩⟨ψ| = 1.
For detailed discussions on quantum computing and information, you can selectively refer to ([4], [5]).

## 3. Previous Work

The two important features of Huffman coding are instantaneousness and variable length. These features are difficult to generalize to the quantum case [6]. The quantum coding field is pioneered by Schumacher works on block coding scheme and the discussions of this coding theorem proofs ([7], [8]). To encode N signals *sequentially*, it requires $O(N^3)$ computational steps. The encoding and decoding processes are far from instantaneous. Moreover, the lengths of all the codewords are the same.

Recently, a successful implementation for a Huffman-coding-inspired scheme for the storage of quantum information is published [6]. This scheme is highly efficient. The encoding and decoding processes of N quantum signals can be done *in parallel* with depth polynomial in (log N), but if parallel machines are unavailable, this encoding scheme will still take only $O(N(\log N)^a)$ computational steps for a sequential implementation.

## 4. Quantum-inspired Huffman Coding

This work began with an observation on a homework! The third author was solving some exercises on Huffman codes on his self study on Information Theory and Compression. To summarize his answers, he constructed a table of three columns: number of symbols, the symbols, and the corresponding Huffman codes. He got a table like the following table, where *n* is the number of symbols descendently ordered according to frequency of occurrence and counting only those symbols with non-zero frequency.

| *n* | symbols | Huffman codes |
|---|---|---|
| 2 | 0, 1 | 0, 1 |
| 3 | 0, 1, 2 | 0, 10, 11 |
| 4 | 0, 1, 2, 3 | 00, 01, 10, 11 |
| 5 | 0, 1, 2, 3, 4 | 00, 01, 10, 110, 111 |
| 6 | 0, 1, 2, 3, 4, 5 | 00, 01, 100, 101, 110, 111 |
| 7 | 0, 1, 2, 3, 4, 5, 6 | 00, 010, 011, 100, 101, 110, 111 |
| 8 | 0, 1, 2, 3, 4, 5, 6, 7 | 000, 001, 010, 011, 100, 101, 110, 111 |
| 9 | 0, 1, 2, 3, 4, 5, 6, 7, 8 | 000, 001, 010, 011, 100, 101, 110, 1110, 1111 |

(Table 1: Huffman codes with different number of symbols)

We restrict our discussion to the case where all the symbols have same frequency, or a special case where all the symbols have same frequency except the last two ones. Hence, we can ignore the frequencies. The observation is that at $n = 2^r$, *r* is some integer > 0  i.e. *n* = 2, 4, 8, … and let us identify these values as the *exact two-based values*, we find that the generated sequence corresponds to an enumeration of binary numbers equivalent to the input symbols. While at other values of *n*, we find that the generated sequence is the same as the sequence generated with (*n*-1) symbols, but with replacing the *last small code*, and let us denote it as $c_{sl}$, with two codes ($c_{sl}0$) and ($c_{sl}1$), as shown below.

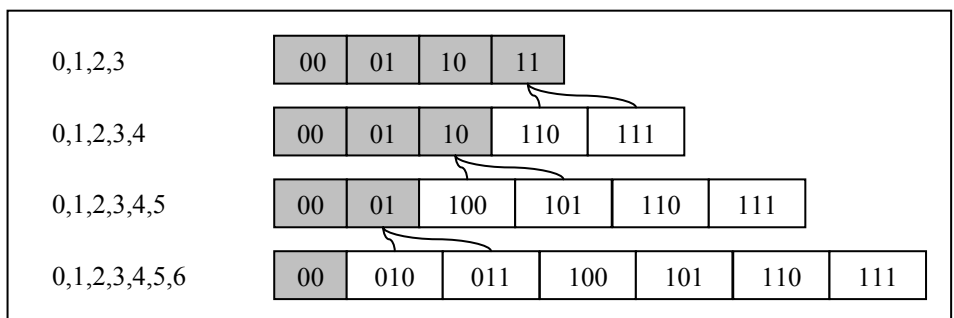

(Figure 3: *last small code* replacement)

We trace how Huffman codes changes while increasing the number of symbols. From now, we shall think of Huffman codes as a sequence corresponding to a certain input sequence. We are eager to find a direct mapping from the input sequence to the Huffman generated sequence. Such a mapping is supposed to be performed in less computational steps.

Now, we have a mapping that seemed to be inspired from studying quantum computing. In order to present this mapping, let us first define some notations:

$lower = \lfloor \lg n \rfloor$, where $\lfloor x \rfloor$ is the floor of $x$
$upper = \lceil \lg n \rceil$, where $\lceil x \rceil$ is the ceil of $x$

It is clear that at the **exact two-based values**, i.e. at $n = 2^r$, $r$ is some integer $> 0$, we find that $lower = upper$ and that is why the generated sequence corresponds to an enumeration of binary numbers equivalent to the input symbols. In the other case, the generated sequence differs than the last binary enumeration and the number of differences is determined by

$diff = 2 (n - 2^{lower})$

Also, we have

$c(x)$ denotes the code of symbol $x$
$b_l(x)$ denotes the binary code of number $x$, with **zero left padding** to get binary codes of length $l$

Now, let us analyze (table 1) using these notations in order to capture a relation between the input sequence and the Huffman sequence.

| | | | | | | | |
|---|---|---|---|---|---|---|---|
| **n = 2**, | $lower = 1$, | $upper = 1$, | $diff = 0$ | **n = 3**, | $lower = 1$, | $upper = 2$, | $diff = 2$ |
| | $c(0) = b_1(0)$ | | | | $c(0) = b_1(0)$ | | |
| | $c(1) = b_1(1)$ | | | | $c(1) = b_2(2)$ | | |
| | | | | | $c(2) = b_2(3)$ | | |
| **n = 4**, | $lower = 2$, | $upper = 2$, | $diff = 0$ | **n = 5**, | $lower = 2$, | $upper = 3$, | $diff = 2$ |
| | $c(0) = b_2(0)$ | | | | $c(0) = b_2(0)$ | | |
| | $c(1) = b_2(1)$ | | | | $c(1) = b_2(1)$ | | |
| | $c(2) = b_2(2)$ | | | | $c(2) = b_2(2)$ | | |
| | $c(3) = b_2(3)$ | | | | $c(3) = b_3(6)$ | | |
| | | | | | $c(4) = b_3(7)$ | | |
| **n = 6**, | $lower = 2$, | $upper = 3$, | $diff = 4$ | **n = 7**, | $lower = 2$, | $upper = 3$, | $diff = 6$ |
| | $c(0) = b_2(0)$ | | | | $c(0) = b_2(0)$ | | |
| | $c(1) = b_2(1)$ | | | | $c(1) = b_3(2)$ | | |
| | $c(2) = b_3(4)$ | | | | $c(2) = b_3(3)$ | | |
| | $c(3) = b_3(5)$ | | | | $c(3) = b_3(4)$ | | |
| | $c(4) = b_3(6)$ | | | | $c(4) = b_3(5)$ | | |
| | $c(5) = b_3(7)$ | | | | $c(5) = b_3(6)$ | | |
| | | | | | $c(6) = b_3(7)$ | | |
| **n = 8**, | $lower = 3$, | $upper = 3$, | $diff = 0$ | **n = 9**, | $lower = 3$, | $upper = 4$, | $diff = 2$ |
| | $c(0) = b_3(0)$ | | | | $c(0) = b_3(0)$ | | |
| | $c(1) = b_3(1)$ | | | | $c(1) = b_3(1)$ | | |
| | $c(2) = b_3(2)$ | | | | $c(2) = b_3(2)$ | | |
| | $c(3) = b_3(3)$ | | | | $c(3) = b_3(3)$ | | |
| | $c(4) = b_3(4)$ | | | | $c(4) = b_3(4)$ | | |
| | $c(5) = b_3(5)$ | | | | $c(5) = b_3(5)$ | | |
| | $c(6) = b_3(6)$ | | | | $c(6) = b_3(6)$ | | |
| | $c(7) = b_3(7)$ | | | | $c(7) = b_4(14)$ | | |
| | | | | | $c(8) = b_4(15)$ | | |

(Table 2: the relation between the input sequence and the Huffman sequence)

Quantum-inspired Huffman coding has a different approach than the traditional Huffman coding, that is instead of constructing a tree and then traversing it, we easily calculate a quantum system state by summing the multiplications of each input times its corresponding output, using **bra-ket** notation. For instance, for 5 symbols, the system state is

$|0\rangle \langle 0| + |1\rangle \langle 1| + |2\rangle \langle 2| + |3\rangle \langle 6| + |4\rangle \langle 7|$

Coding, i.e. getting the code for a certain input symbol, is done by simply multiply the transpose of the matrix representation of the system state times the matrix representation of the state of that input.

**Calculating the system state**
  Step 0: Input the number of symbols, i.e. $n$
  Step 1: calculate parameters
    $lower := \lfloor \lg n \rfloor$
    $upper := \lceil \lg n \rceil$
    $diff := 2(n - 2^{lower})$

Step 2: Initialize system state as a zero matrix
  *state* := 0
Step 3: If *lower* = *upper* then do
  Step 3.1: for *counter* = 0 to (*n* – 1)
    Step 3.1.1: *state* += |*counter*> <*counter*|
  End for
Else do
  Step 3.2: for *counter* = 0 to (*n* – 1 – *diff*)
    Step 3.2.1: *state* += |*counter*> <*counter*|
  End for

  Step 3.3: for *counter* = (*n* – *diff*) to (*n* – 1)
    Step 3.3.1: *state* += |*counter*> <$2^{upper}$ + *counter* – *n* |
  End for
End if

Note that when *lower* = *upper* , the system state is just a matrix of size (*lg n* × *lg n*). But in the other case, where *lower* ≠ *upper*, the system state is the sum of two matrices , *state*$_1$ and *state*$_2$ say, of sizes (*upper* × *lower*)  and (*upper*×*upper*) respectively. Hence, we require two registers to store the two matrices.

**Coding**
Step 0: Input the number of symbols, i.e. *n* and the system state, i.e. *state*
Step 1: calculate parameters
  *lower* := ⌊ *lg n* ⌋
  *upper* := ⌈ *lg n* ⌉
  *diff* := 2 (*n* - $2^{lower}$ )
Step 2: Input the number of the symbol to be decoded, say *numb*, from 0 to (*n* – 1)
Step 3: If *lower* = *upper* then do
  Step 3.1: *result* := |*numb*> *state*
Else do
  Step 3.2: if *numb* ≤ (*n* – 1 – *diff*) then do
    Step 3.2.1: *result* := |*numb*> *state*$_1^t$     where *state*$_1$ is the matrix stored in the first register
  Else do
    Step 3.2.2: *result* := |*numb*> *state*$_2^t$     where *state*$_1$ is the matrix stored in the second register
  End if
End if

**Example**: For 5 symbols.
  *n* :=5
  *lower* := ⌊ *lg n* ⌋ = 2
  *upper* := ⌈ *lg n* ⌉ =3
  *diff* := 2 (*n* - $2^{lower}$ ) = 2 (5 – 4) = 2

The system state is
$$|0> <0| \;\;+\;\; |1> <1| \;\;+\;\; |2> <2| \;\;+\;\; |3> <6| \;\;+\;\; |4> <7|$$
or precisely,
$$|000> <00| \;\;+\;\; |001> <01| \;\;+\;\; |010> <10| \;\;+\;\; |011> <110| \;\;+\;\; |100> <111|$$

In Matrix representation it is

$$\begin{pmatrix}1\\0\\0\\0\\0\\0\\0\\0\end{pmatrix}(1\;\;0\;\;0\;\;0) + \begin{pmatrix}0\\1\\0\\0\\0\\0\\0\\0\end{pmatrix}(0\;\;1\;\;0\;\;0) + \begin{pmatrix}0\\0\\1\\0\\0\\0\\0\\0\end{pmatrix}(0\;\;0\;\;1\;\;0) + \begin{pmatrix}0\\0\\0\\1\\0\\0\\0\\0\end{pmatrix}(0\;\;0\;\;0\;\;0\;\;0\;\;0\;\;1\;\;0) + \begin{pmatrix}0\\0\\0\\0\\1\\0\\0\\0\end{pmatrix}(0\;\;0\;\;0\;\;0\;\;0\;\;0\;\;0\;\;1)$$

which is summed up to

$$\begin{pmatrix} 1 & 0 & 0 & 0 \\ 0 & 1 & 0 & 0 \\ 0 & 0 & 1 & 0 \\ 0 & 0 & 0 & 0 \\ 0 & 0 & 0 & 0 \\ 0 & 0 & 0 & 0 \\ 0 & 0 & 0 & 0 \\ 0 & 0 & 0 & 0 \end{pmatrix} + \begin{pmatrix} 0 & 0 & 0 & 0 & 0 & 0 & 0 & 0 \\ 0 & 0 & 0 & 0 & 0 & 0 & 0 & 0 \\ 0 & 0 & 0 & 0 & 0 & 0 & 0 & 0 \\ 0 & 0 & 0 & 0 & 0 & 0 & 1 & 0 \\ 0 & 0 & 0 & 0 & 0 & 0 & 0 & 1 \\ 0 & 0 & 0 & 0 & 0 & 0 & 0 & 0 \\ 0 & 0 & 0 & 0 & 0 & 0 & 0 & 0 \\ 0 & 0 & 0 & 0 & 0 & 0 & 0 & 0 \end{pmatrix}$$

We try now to get the code of an input of number lass then or equal ($n - 1 - diff$) i.e. 2, say the third symbol that has number 2. Then we try to get the code of another input of number grater than ($n - 1 - diff$), say the fourth symbol that has number 3.

To get the code of the third symbol, simply multiply the transpose of the first matrix by |2>

$$\begin{pmatrix} 1 & 0 & 0 & 0 & 0 & 0 & 0 & 0 \\ 0 & 1 & 0 & 0 & 0 & 0 & 0 & 0 \\ 0 & 0 & 1 & 0 & 0 & 0 & 0 & 0 \\ 0 & 0 & 0 & 0 & 0 & 0 & 0 & 0 \end{pmatrix} \begin{pmatrix} 0 \\ 0 \\ 1 \\ 0 \\ 0 \\ 0 \\ 0 \\ 0 \end{pmatrix} = \begin{pmatrix} 0 \\ 0 \\ 1 \\ 0 \end{pmatrix} = |2\rangle = |10\rangle$$

To get the code of the fourth symbol, simply multiply the transpose of the second matrix by |3>

$$\begin{pmatrix} 0 & 0 & 0 & 0 & 0 & 0 & 0 & 0 \\ 0 & 0 & 0 & 0 & 0 & 0 & 0 & 0 \\ 0 & 0 & 0 & 0 & 0 & 0 & 0 & 0 \\ 0 & 0 & 0 & 0 & 0 & 0 & 0 & 0 \\ 0 & 0 & 0 & 0 & 0 & 0 & 0 & 0 \\ 0 & 0 & 0 & 0 & 0 & 0 & 0 & 0 \\ 0 & 0 & 0 & 1 & 0 & 0 & 0 & 0 \\ 0 & 0 & 0 & 0 & 1 & 0 & 0 & 0 \end{pmatrix} \begin{pmatrix} 0 \\ 0 \\ 0 \\ 1 \\ 0 \\ 0 \\ 0 \\ 0 \end{pmatrix} = \begin{pmatrix} 0 \\ 0 \\ 0 \\ 0 \\ 0 \\ 0 \\ 1 \\ 0 \end{pmatrix} = |6\rangle = |110\rangle$$

Now, let us present a simpler scheme of calculating the Huffman code corresponding to a certain input.

**Direct Mapped Huffman Codes**
    Step 0: Input the number of symbols, i.e. *n*
    Step 1: calculate parameters
        $lower := \lfloor lg\ n \rfloor$
        $upper := \lceil lg\ n \rceil$
        $diff := 2\ (n - 2^{lower})$
    Step 2: Input the number of the symbol to be decoded, say *numb*, from 0 to ($n - 1$)
    Step 3: If *lower* = *upper* Or *numb* ≤ ($n - 1 - diff$) then do
        Step 3.1: *result* := binary (*numb*, *lower*)
    Else do
        Step 3.2: *result* := binary ($2^{upper} + numb - n$, *upper*)
    End if

Where binary (*numb*, *len*) denotes the binary number of number *numb*, with *zero left padding* to get a binary code of length *len*. We are conceived that calculating a Huffman code is O(1) now, as all computational steps are built-in with most today's computer processors, including converting an integer to its corresponding binary number and padding. But for scientific precision, this scheme takes O(*lg n*) for calculating a Huffman code considering the time taken to convert the input integer to its corresponding binary number and padding.

## 5. Results and Conclusions

The traditional Huffman coding has two procedures : constructing a tree and then traversing it. Constructing the tree is $O(n^2)$, yet it can be optimized using appropriate data structures and programming algorithms. Traversing the tree is $O(n)$.

Our algorithm, Quantum-inspired Huffman coding, also has two procedures : calculating a quantum system state and then multiplying it by the inputs. Calculating the quantum system state is $O(n (\lg n)^2)$. Multiplying the quantum system state by an input is $O((\lg n)^2)$.

It is clear that $O(n (\lg n)^2)$ is smaller than $O(n^2)$ and that $O((\lg n)^2)$ is smaller than $O(n)$ for all values of *n* greater than 16, as $(\lg n)^2 = n$ at *n* =16.

Although our work that is titled "*Quantum-inspired Huffman coding*" may be similar to the previous work [6] that is titled "*A Quantum Analog of Huffman Coding*", we should note here that we reach this idea and scheme independently. Another clear thing to note is the difference in the way of merging the two technologies of *Quantum Computing* and *Huffman coding*.

Finally, we present an unprecedented scheme for direct mapped Huffman codes that is $O(\lg n)$.

## 6. Future Work

We should not forget that we have restricted our discussion to the case where all the symbols have same frequency, and a special case where all the symbols have same frequency except the last two ones. The future work may be the try to generalize the scheme presented here to be applied to the general case where the frequencies are different.

Another future work is application of this scheme in place of the Traditional Huffman coding or its adaptive form in some well known combined techniques as in ([9], [10], [11])